\documentclass[10pt,preprint]{iopart}
\usepackage{iopams}
\usepackage{graphicx}
\usepackage[english]{babel}

\begin{document}
\title{THz-driven demagnetization with Perpendicular Magnetic Anisotropy: Towards ultrafast ballistic switching}
\author{Debanjan Polley$^1$, Matteo Pancaldi$^2$, Matthias Hudl$^1$, Paolo Vavassori$^{2,3}$, Sergei Urazhdin$^4$, Stefano Bonetti$^1$}
\address{$^1$ Department of Physics, Stockholm University, 106 91 Stockholm, Sweden}
\address{$^2$ CIC nanoGUNE, E-20018 Donostia-San Sebastian, Spain}
\address{$^3$ IKERBASQUE, Basque Foundation for Science, E-40013 Bilbao, Spain}
\address{$^4$ Emory University, Atlanta, GA 30322, USA}
\ead{stefano.bonetti@fysik.su.se}

\begin{abstract}
We study THz-driven spin dynamics in thin CoPt films with perpendicular magnetic anisotropy. Femtosecond magneto-optical Kerr effect measurements show that demagnetization amplitude of about $1\%$ can be achieved with a peak THz electric field of $300$~kV/cm, and a corresponding peak magnetic field of $0.1$~T. The effect is more than an order of magnitude larger than observed in samples with easy-plane anisotropy irradiated with the same field strength. We also utilize finite-element simulations to design a meta-material structure that can enhance the THz magnetic field by more than an order of magnitude, over an area of several tens of square micrometers. Magnetic fields exceeding $1$~Tesla, generated in such meta-materials with the available laser-based THz sources, are expected to produce full magnetization reversal via ultrafast ballistic precession driven by the THz radiation. Our results demonstrate the possibility of table-top ultrafast magnetization reversal induced by THz radiation.
\end{abstract}

\submitto{\JPCM}
\maketitle
\ioptwocol

\section{Introduction}

The pioneering observation of ultrafast demagnetization induced in thin films by femtosecond laser pulses~\cite{Beaurepaire1996} has ushered in the  field of ultrafast magnetism. However, despite the ongoing intense experimental and theoretical research~\cite{koopmans2000ultrafast, koopmans2005unifying, RasingPRB2010, PhysRevB.78.174422, Munzenberg:NatureMaterials:2010, koopmans2005unifying, Stamm:NatureMaterials:2007, dalla2007influence, PhysRevB.78.174422, malinowski2008control,  Koopmans:NatureMaterials:2009,Boeglin:Nature:,kirilyuk2010ultrafast, battiato2010superdiffusive, radu2011transient, mathias2012probing, guidoni2002magneto, ostler2012ultrafast, mentink2012ultrafast, schellekens2013comparing, carva2013ab, Turgut:PhysicalReviewLetters:2013, bergeard2014ultrafast, Mangin:NatureMaterials:2014, lambert2014all}, fundamental microscopic understanding of ultrafast demagnetization is still lacking. Most importantly, the mechanism responsible for the fast dissipation of spin angular momentum, which is necessarily involved in the demagnetization process, has not yet been established. The problem in identifying this mechanism is associated mostly with the difficulty of modeling the non-equilibrium states created by the femtosecond pulses of near-infrared (NIR) radiation typically utilized to induce ultrafast demagnetization.

Intense terahertz (THz) radiation was recently shown to provide an alternative approach to inducing ultrafast demagnetization in thin metallic films \cite{StefanoPRL2016, Hauri2016, shalaby2016simultaneous}. There are several substantial differences between the NIR and the THz regimes. In the NIR-induced demagnetization, the electronic temperature rises above $1000$~K \cite{rhie2003femtosecond}, and the subsequent cooling of the electron gas dominates over the individual scattering events~\cite{StefanoPRL2016}. In contrast, in the THz-induced demagnetization, the electronic temperature is only slightly increased, so the individual electron scattering events become more important for the subsequent relaxation. In particular, it was shown that the time scale for the inelastic spin scattering is by an order of magnitude shorter in the THz regime than in the NIR regime ~\cite{StefanoPRL2016}. It was also shown that the crystalline order of the magnetic material determines the inelastic spin scattering rate in the THz regime, while it is not important fo the spin relaxation in the NIR-regime~\cite{StefanoPRL2016}. Finally, the large-amplitude magnetic field produced by the THz radiation in metallic ferromagnets can induce coherent magnetization precession. However, this effect has only been demonstrated in the low-amplitude limit only on a few magnetic materials, all with easy-plane magnetic anisotropy~\cite{StefanoPRL2016, Hauri2016, shalaby2016simultaneous, Hauri2013nature}.

In this article, we present measurements and analysis of the effects of THz radiation on thin metallic films with perpendicular magnetic anisotropy (PMA). Ferromagnetic thin films with PMA have been studied for applications in nanoscale spintronic devices such as spin-transfer-torque magnetic random access memories (STT-MRAMs)~\cite{KentSpintronics2010,PolleyPMATHz2015}. All-optical switching of magnetization was recently demonstrated in CoPt multilayers with PMA~\cite{lambert2014all}. Here, we experimentally investigate the ultrafast spin dynamics of a similar multilayer driven by strong THz radiation. In addition, we design a meta-material structure that can provide a strong enhancement of the THz radiation, enabling generation of sufficiently large magnetic fields to induce coherent precessional (ballistic) magnetization reversal~\cite{stohr2007magnetism}. In this regime, the magnetization is precessionally ``dragged'' by the magnetic field pulse across the magnetic energy maximum, subsequently relaxing to the reversed equilibrium state~\cite{analyticaldamping2003}. The precessional energy barrier is lowered by the PMA, reducing the THz field amplitude required for the ballistic magnetization switching.

\section{Sample and experimental geometry}
\begin{figure*}[t]
\centering
\includegraphics[width=0.8\textwidth]{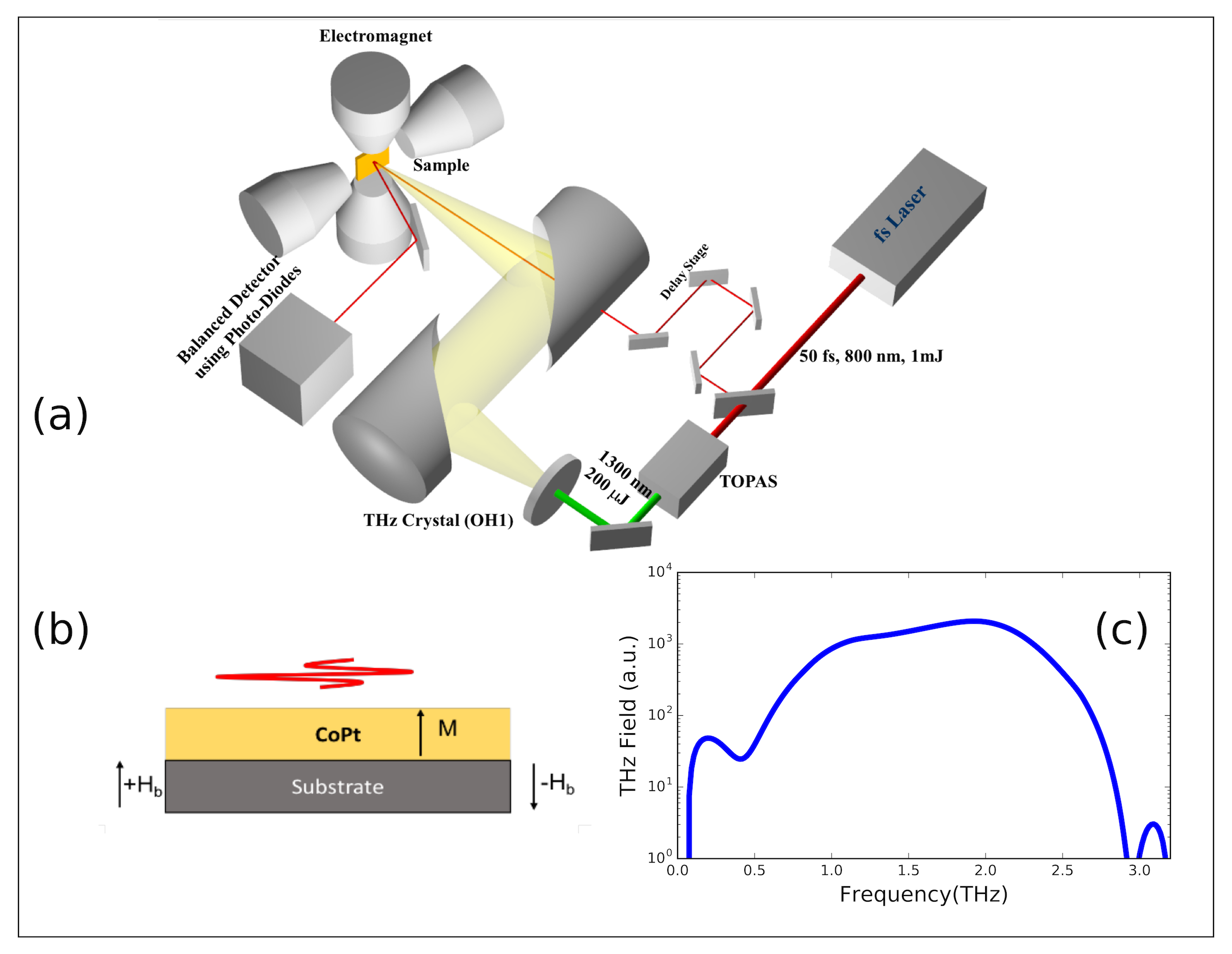}
\caption{(a) Schematic of the THz pump-800 nm probe setup. The electromagnet is capable of three-dimensional rotation of the static magnetic field. (b) Schematic of the sample and the measurement geometry:  The magnetization $\mathbf{M}$ is aligned by the static magnetic field $\mathbf{H_b}$ applied perpendicular to film plane. Both the electric and the magnetic field components of the normally incident THz radiation are in the film plane, perpendicular to $\mathbf{M}$. (c) Frequency spectrum of the THz fields used in our experiments, calculated as the Fourier transform of the electro-optical sampling signal from GaP.}
\label{F1}
\end{figure*}

The studied ferromagnetic thin film with perpendicular magnetic anisotropy (PMA) had the multilayer Ta(4) / Pt(5) / Co(1) / Pt(0.5) / Co(1) / Cu(0.4) / Ta(2), with the thicknesses of layers given in nanometers. The sample was grown on a $500~\mu$m-thick oxidized silicon substrate. The schematic of the experimental geometry is shown in Fig.~\ref{F1}(a). The sample was mounted in an electromagnet that produced magnetic field of up to $100$~mT normal to the film plane.

The strong THz radiation was generated by optical rectification, in a OH1 \cite{OH12008} crystal (Rainbow Photonics), of a $60$~fs-long, 200 $\mu$J laser pulse centered at the wavelength of $1300$~nm. The optical pulse was generated via optical parametric amplification from a $50$~fs-long, $1$~mJ pulse at $800$~nm wavelength, produced by a $1$~kHz regenerative amplifier. The amplitude and the temporal shape of the THz electric field were characterized by electro-optical (EO) sampling in a $100$~$\mu$m-thick GaP crystal (not shown). The maximum peak electric field reach approximately $300$~kV/cm, and the corresponding peak magnetic field reached about $100$~mT in our measurements, as was calculated using Eq.~(1) in Ref.~\cite{hoffmann2011coherent}. The Fourier spectrum of the THz pulses gives the FWHM bandwidth of approximately 1.5 THz, with a 1.4 THz center frequency, as shown in \ref{F1}(c).

The ultrafast spin dynamics in CoPt was driven by the intense THz pulse incident normal to the films surface, as schematically shown in Fig. \ref{F1}(b). A small fraction of the NIR laser light from the regenerative amplifier was used to stroboscopically probe the magnetization dynamics, by utilizing the femtosecond magneto-optical Kerr effect (MOKE). The probe pulse delay, relative to the pump pulse, was controlled with a motorized delay stage.

\section{Spin dynamics in CoPt films with PMA}
\begin{figure*}[t]
\centering
\includegraphics[width=\textwidth]{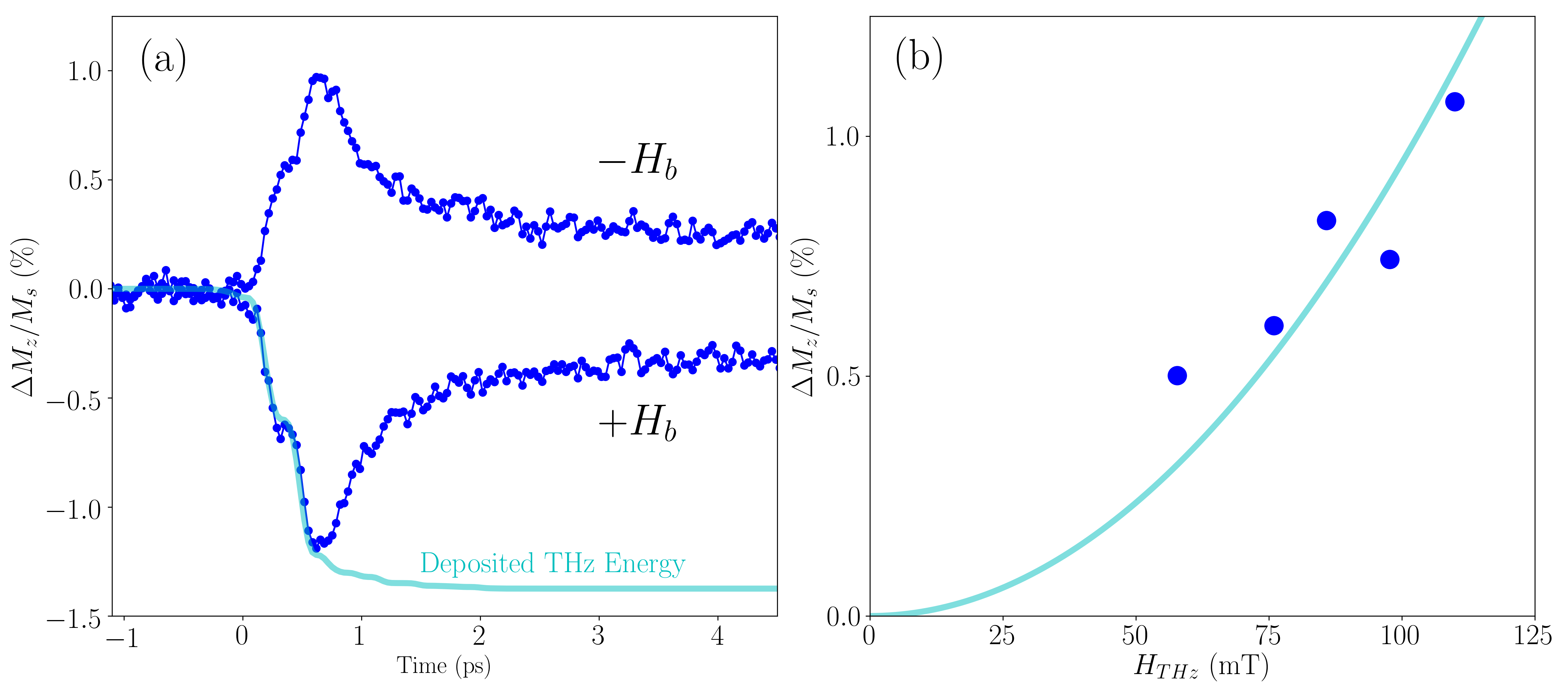}
\caption{(a) Symbols connected by lines: Time-resolved polar MOKE signal, measured using the CoPt film described in the text, for two opposite directions of the static field normal to the film. The signal is normalized to the amplitude of the static MOKE signal. The THz pulse that excites the film arrives at time $0$~ps. Curve: the calculated energy deposited by the THz radiation. (b) (Symbols) Maximum measured demagnetization as a function of THz magnetic field amplitude. The solid line shows the quadratic dependence.}
\label{F3}
\end{figure*}

Fig.~\ref{F3}(a) shows the time-dependent MOKE signal from the CoPt film at the largest THz fields. The signal rapidly increases over the THz pulse duration of about $0.7$~ps, and subsequently slowly decays. This dependence is reversed when the direction of the static magnetic field $|H_b|=80$ mT is reversed, confirming its magnetic origin. The observed time-dependent MOKE signal can therefore be identified as ultrafast demagnetization. A curve in Fig.~\ref{F3}(a) shows the THz energy deposited in the sample, calculated as he time-integral of the THz field measured independently by the EO sampling. The two-step structure, observed in the initial rapid increase of the MOKE signal, is well-reproduced by the step structure of the deposited THz energy, 
 in agreement with the previous work~\cite{StefanoPRL2016}. In that work, the two-step structure was not as well resolved, due to the lower temporal resolution of $100$~fs. The higher temporal resolution (approximately 50 fs) of our measurement clearly resolves this two-step structure, and confirms that the spin-lattice scattering occurs on the same time scale of $\approx30$~fs as the spin-conserving electronic scattering, as was originally demonstrated in Ref.~\cite{StefanoPRL2016}. Figure~\ref{F3}(b) shows that the demagnetization amplitude is proportional to the square of THz field amplitude, also in agreement with the previous work.

\begin{figure}[b]
\centering
\includegraphics[width=\columnwidth]{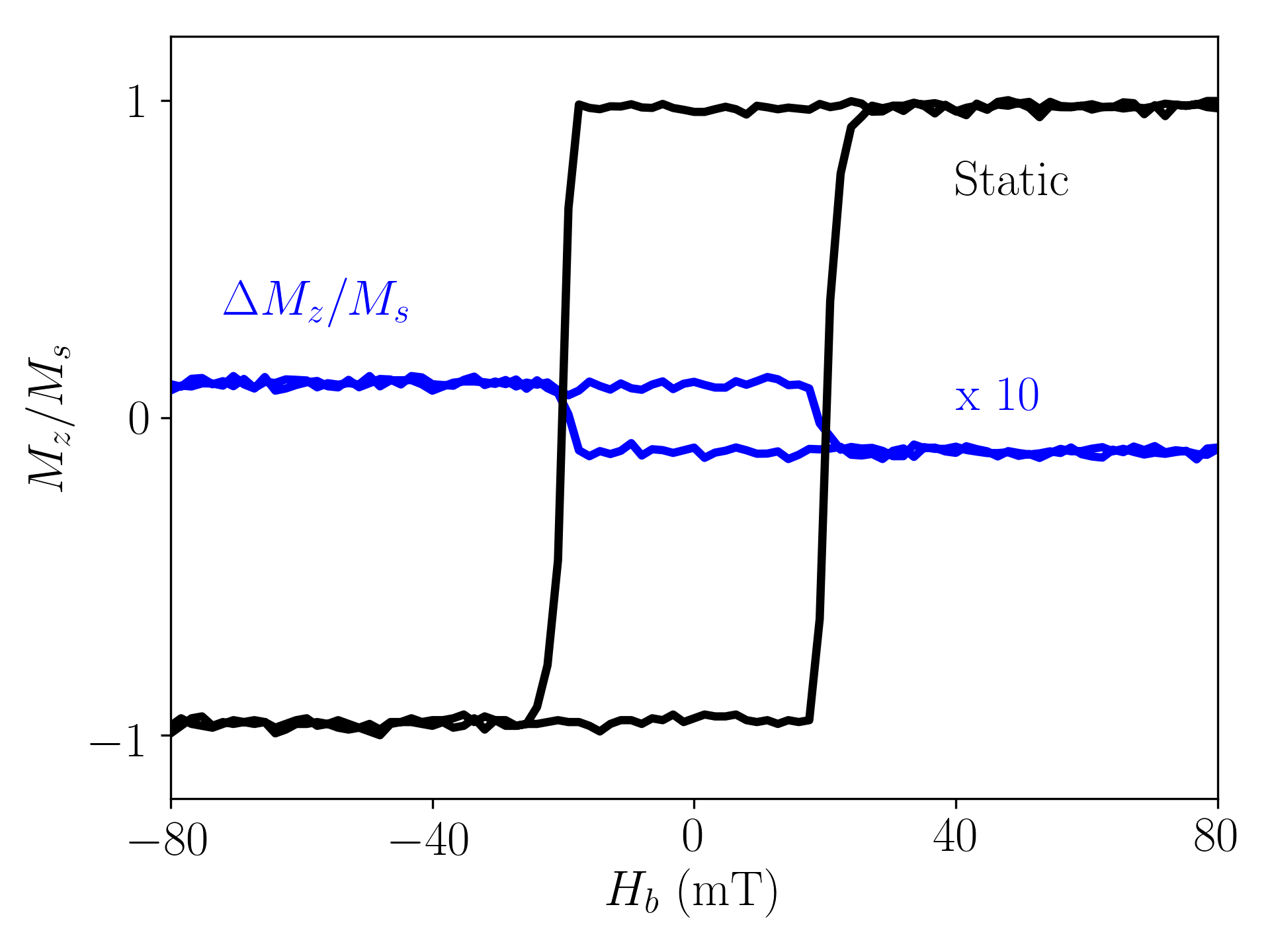}
\caption{Normalized magnetization loops measured by polar MOKE signal, measured using the CoPt film described in the text. The black curves show the magnetization loop obtained without the THz field, while the blue lines show the \emph{differential} magnetization loop obtained with the THz field, at the delay $0.7$~ps corresponding to the maximum THZ-induced demagnetization.}
\label{F2}
\end{figure}

In contrast to Ref.~\cite{StefanoPRL2016}, in this experiment we could not resolve the coherent response caused by the torque exerted on the magnetization by the magnetic field of the THz radiation. This can be explained by considering the difference between the two geometries: in an in-plane magnetized sample, where the magnetization lies at a polar angle $\theta_0\approx0$ before the arrival of the THz pulse, the polar MOKE signal probing the out-of-plane component of the magnetization varies as $\sin\theta\sim\theta$ for small deviations from equilibrium. In contrast, for the perpendicularly magnetized sample studied in our experiment ($\theta_0\approx90$ deg), at small variations of the magnetization angle induced by the THz pulse, the polar MOKE signal remains close to its maximum value, and varies as $\sim(\theta-\theta_0)^2$. We note that the incoherent (demagnetization) and coherent (precession) effects of the THz field are generally expected to always coexist, but the geometry of the measurement must be specially tailored to disentangle these two contributions.

In Fig.~\ref{F2}, we compare the static magnetic hysteresis loop of the CoPt film, measured without the THz radiation (black curve), with the hysteresis loop acquired at the probe delay of $0.7$~ps after the THz pulse, when the film exhibits the largest demagnetization (see Fig.~\ref{F3}(a)). The coercivity of the film is approximately $20$~mT in both measurements, unaffected by the THz excitation, indicating that the effects of heating are not important.

As evidenced from Figs.~\ref{F3}(a) and \ref{F2}, the maximum observed demagnetization is greater than 1\% for the studied CoPt film with PMA. This is remarkable, as the demagnetization is an order of magnitude larger than in the previously studied CoFeB films with easy-plane anisotropy, at similar THz fields strength. We note that this observation is consistent with the previously suggested model of defect-driven spin-lattice scattering as the driving force behind THz-induced ultrafast demagnetization. Indeed, the ultrathin Pt layers in the CoPt multilayer likely act as efficient scatters characterized by a large spin-orbit interaction, resulting in a much higher spin-flip rate of electrons in CoPt than in CoFeB. We leave more detailed analysis of this observation to future studies.

\begin{figure*}[t]
\centering
\includegraphics[width=0.9\textwidth]{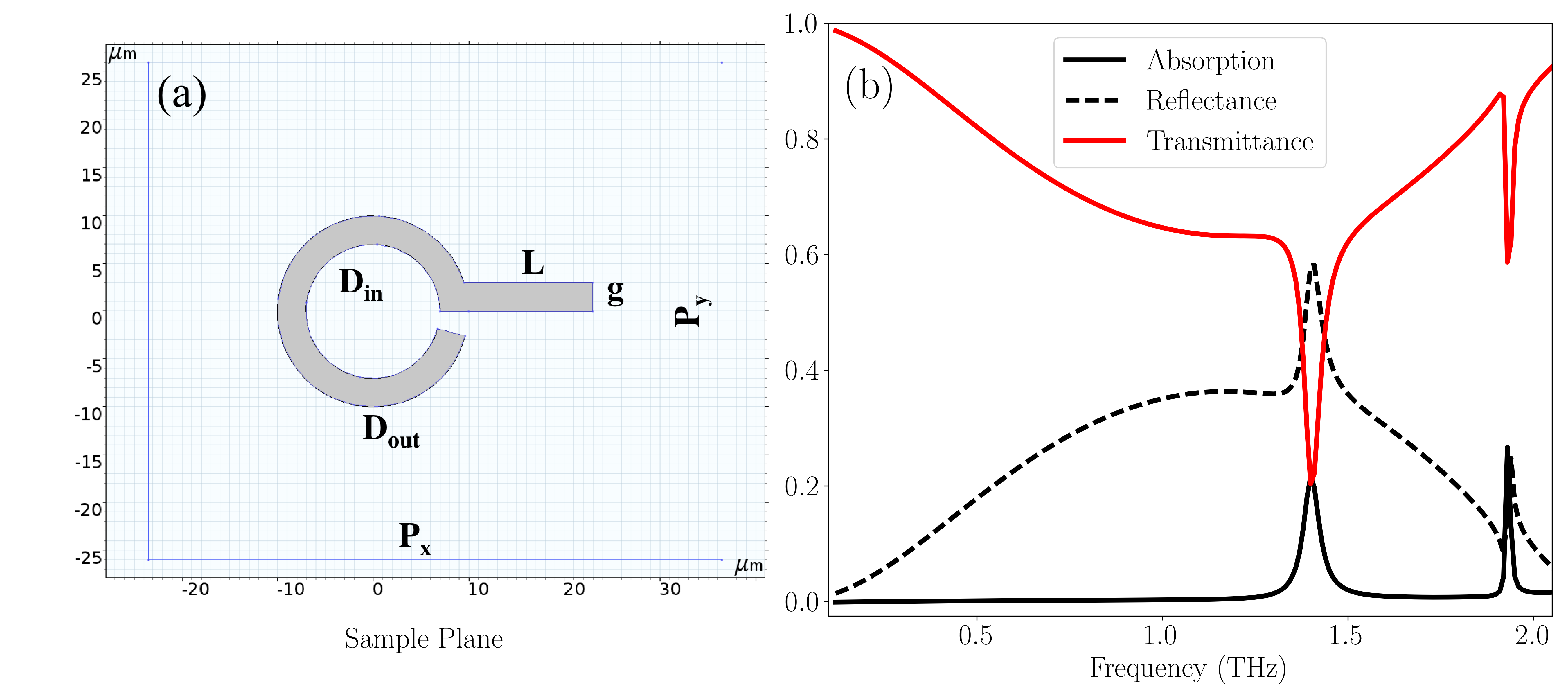}
\caption{ (a) Schematic of the unit cell of the studied meta-material. The parameters are $P_{x}$=$60$ $\mu$m, $P_{y}=52$ $\mu$m, $D_{in}=14$ $\mu$m, $D_{out}=20$ $\mu$m, $L=16$ $\mu$m and $g=3$ $\mu$m, following the nomenclature described in the text, (b) Transmittance, absorption and reflectance of the THz MM structure in the frequency range of $0.1-2.5$~THz.}
\label{F4}
\end{figure*}

For the rest of this work, we are only interested in the fact that CoPt, one of the few materials where all-optical magnetization switching in the NIR has been demonstrated~\cite{lambert2014all}, also exhibits enhanced THz-driven ultrafast demagnetization. To become relevant to the magnetization switching, the effects of the electromagnetic field need to be further enhanced by two orders of magnitude. Here, we propose a two-fold strategy. First, one can use larger THz fields. A three-fold increase from 300 kV/cm to 1 MV/cm (achievable with the present-day lasers and demonstrated in several laboratories), would already lead to the 9 times larger demagnetization, exceeding 10\% in total. Second, one can design  THz meta-materials to achieve near-field enhancement of the THz radiation by another order of magnitude. In the next section, we propose a design of the meta-material structure that can accomplish this goal.

\section{THz meta-material structure}
Metamaterials (MMs) are commonly formed by sub-wavelength conductive elements patterned in a periodic fashion, and usually deposited on a dielectric substrate. They can be used to tailor the electromagnetic response that cannot be realized in naturally occurring materials~\cite{chen2008THzactiveMM,pancaldiTHzantenna2017}. The structure can be either resonant or non-resonant \cite{chen2008THzactiveMM,PolleyTHzanti}, and the resonance frequency can be optimized by modifying the structure and the geometry of the MM. Here, we have numerically studied the local THz electric and magnetic field enhancement in a resonant MM structure using the finite-element method implemented in the COMSOL Multiphysics software.

The unit cell of the studied THz MM structure consists of a $200$~nm-thick gold Au split-ring-resonator, on top of a $30$~$\mu$m-thick glass substrate, as schematically shown in Fig.~\ref{F4}(a). The structure also includes air blocks with thickness $\lambda$ - the wavelength of the corresponding THz radiation - on top of the gold structure and at the bottom of the glass substrate. Additionally, to avoid spurious reflections of the electromagnetic waves, we have included perfectly matched layers with thickness $\lambda/5$ at the top and the bottom boundaries of the simulation region. Two identical ports are used at the end of each air block, one to launch the THz plane wave, the other to measure the THz radiation transmitted through the structure. The length $(P_{x})$ and width $(P_{y})$ of the unit cell are 60 $\mu$m and 52 $\mu$m, respectively. The inner diameter $D_{\rm in} $ of the Au ring resonator is 14 $\mu$m, and its outer diameter $D_{\rm out} $ is 20 $\mu$m. A 15-degree cutout in the resonator ring is designed to produce the split-ring structure. An arm with the length $L=16$~$\mu$m and width $g= (D_{\rm out}-D_{\rm in})/2=3$ $\mu$m, is added to the Au split-ring to introduce an asymmetry in the structure, such that the magnetic field of the incident THz radiation, perpendicular to the arm, can efficiently couple with the resonator.

\begin{figure*}[t]
\centering
\includegraphics[width=\textwidth]{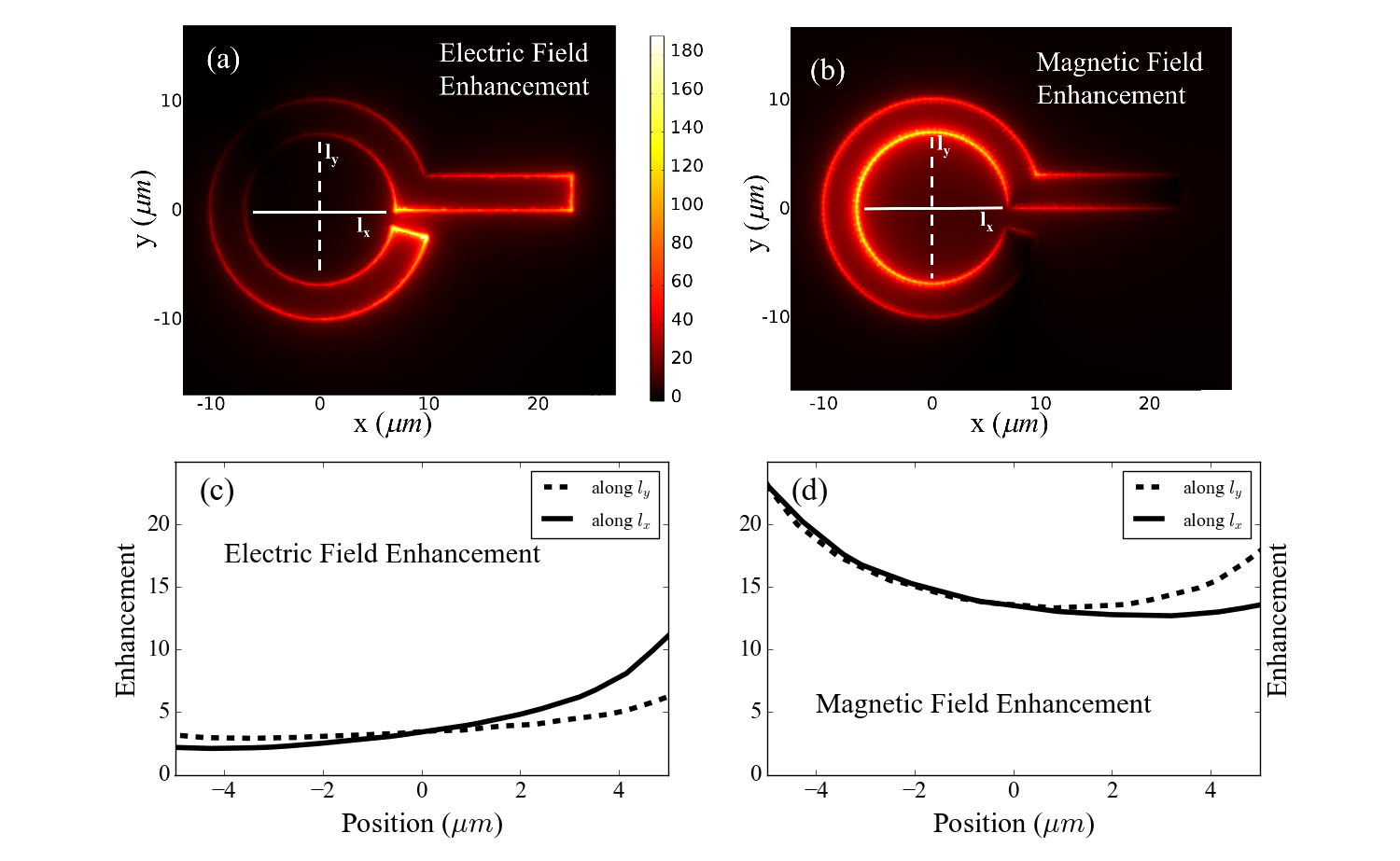}
\caption{Upper panels: enhancement map for (a) THz electric and (b) THz magnetic fields at the MM surface, at $1.4$~THz. Lower panels: enhancement of the THz electric (c) and magnetic (d) fields in the direction $l_{x}$ parallel to the arm (solid line), and the direction $l_{y}$ perpendicular to the arm (dashed line). The THz radiation, impinging along the normal to the simulation surface, is linearly polarized, with the electric field directed parallel to the arm, and the magnetic field orthogonal to it.}
\label{F5}
\end{figure*}

The parameters of the resonator were optimized to obtain the resonant frequency of 1.4 THz, close to the center frequency of the THz radiation in our measurements, as shown in Fig. \ref{F1}(c). To account for the periodicity of the meta-material, perfect magnetic conductor (PMC) and perfect electric conductor (PEC) boundary conditions are applied to the surfaces perpendicular to the $x$ and $y$ directions, respectively.

To perform finite-element simulations, tetrahedral mesh was created with the mesh size of $g/5.5\approx0.5$ $\mu$m and the minimum mesh size of $g/15\approx0.2$ $\mu$m. The substrate and the air blocks are divided into tetrahedral meshes with a maximum mesh size of $\lambda /(20n) \sim $ and $\lambda /15 $, respectively, and the minimum mesh size fixed at $g/5.5\approx0.5$ $\mu$m. Here, $n$ is the real refractive index of the substrate. The metallic Au layer is modeled as a lossy metal with the frequency independent conductivity of gold, $\sigma = 4.09 \cdot 10^7$ Sm$^{-1}$~\cite{AuTHzValue}. Experimentally determined optical constants are used in modeling the glass substrate~\cite{thzglass}.

The simulations are performed at frequencies in the range of 0.1 to 2.5 THz in steps of 0.01 THz. The transmittance, the reflectance, and the absorption of the structure all exhibit sharp features near 1.4 THz due to the presence of the MM structure, as shown in Fig. 4b. In particular, the absorption exhibits a peak with the maximum value of $\sim20\% $, the reflectance peaks at $\sim 60\%$, and the transmittance dips to $\sim 20\% $ at 1.4 THz. The low transmittance at 1.4 THz provides a strong indication that the structure should  induce THz field enhancement. Both electric and magnetic fields of the THz radiation are generally expected to be locally enhanced. THz-induced ballistic magnetization reversal requires enhancement of the THz magnetic field, while the electric field enhancement is likely unimportant for this process. We achieve this by using the arm structure in the MM: its coupling with the incident THz field induces a current that screens its magnetic field component. This current then flows through the ring, which in turn results in a strong THz magnetic field inside the circular area of the structure.

To estimate the enhancement factor, we used the following approach. In the simulation, the THz electric and magnetic fields incident on the metamaterial are taken as reference, and are assumed to be uniform over the simulation surface. Once the THz electric and magnetic field values around the metamaterials are determined, these calculated values are normalized by the reference fields. The resulting enhancement maps for the electric and the magnetic fields are the plotted in Fig.~\ref{F5}.

The enhancement of the THz electric and magnetic field at 1.4 THz is shown in Fig. \ref{F5}(a) and (b), respectively. The THz magnetic field is strongly enhanced close to the Au ring, and most significantly adjacent to the inner edge of the ring. The field enhancement exhibits relatively minor variations over the entire inner region  of the MM structure with approximately 100 $\mu$m$^{2}$ area, and very uniform over the area in the center of a few micrometers squared. Figures~\ref{F5}(c) and \ref{F5}(d) show the cross-sections indicated in the maps by lines $l_{x}$ and $l_{y}$, respectively. These cross-sections clearly show that inside the circular region, the THz magnetic field is almost uniformly enhanced by approximately a factor of 13, compared to the incident field. The THz electric field is also enhanced, on average by less than a factor of 2, and it is not as uniform. For metallic structures that could be placed within the ring, the THz electric field is not important, as it is screened very efficiently at the surface of the conductor.

This MM structure can therefore provide local THz magnetic fields with the magnitude exceeding $1$~T, using the free-space THz field pf  100 mT utilized in this work to demagnetize the CoPt films. For free-space THz fields of the order of $0.3-0.6$ T recently demonstrated by several groups, local THz magnetic fields approaching 10 T should be achievable with this MM structure. In both cases, full demagnetization of a micro- or nanometer-scale patterned sample enclosed in the meta-material become possible, thanks to the quadratic dependence of the demagnetization on the THz field amplitude.

As noted above, one of the distinguishing characteristics of the THz radiation, compared to the NIR radiation, is that it can drive substantial magnetization precession via the torque induced by its magnetic field component. This could open up for the realization, in a table-top experiment, of ultrafast precessional (ballistic) magnetization switching, which is believed to provide the fastest possible  magnetization reversal \cite{tudosa2004ultimate}. The feasibility of the ballistic magnetization reversal by the THz radiation has been demonstrated with THz magnetic fields of the order of 1 T generated by a relativistic electron beam in a two-mile long linear accelerator  \cite{tudosa2004ultimate,gamble2009electric}, in a PMA sample similar to the one investigated here. However, because of the complexity associated with such a large-scale experiment, direct real-time observation of such ballistic switching has not yet been achieved. By producing similar magnitudes of the THz fields with a combination of laser-based setup and meta-materials, time-resolved investigations of the THz-induced magnetization reversal with femtosecond resolution may now become possible.

\section{Conclusion}

We utilized the femtosecond magneto-optical Kerr effect to probe the ultrafast spin dynamics induced by strong THz fields in a CoPt multilayer thin film with perpendicular magnetic anisotropy. We observed a ten-fold increase in the demagnetization amplitude, as compared to the samples with easy-plane anisotropy at the same THz field strength. In addition, we have utilized the finite-element method to numerically calculate the enhancement of the magnetic field of the THz radiation in a modified split-ring-resonator meta-material structure. We found that the enhancement is greater than a factor of 10, and is uniform over an area of the order of $100$~$\mu$m$^2$. We predict that the strong THz radiation enhanced by such a meta-material structure can be sufficient to induce ballistic magnetization reversal in magnetic films with perpendicular magnetic anisotropy, where the anisotropy barrier for the ballistic switching is lower than in films with in-plane anisotropy. While the regime of large THz fields is still largely unexplored, we can generally expect a substantial precessional character for such reversal, due to the torque exerted by the THz magnetic field on the magnetization. This regime is qualitatively different from NIR-induced all-optical switching, and can lead to the table-top observation of ultrafast ballistic magnetization switching, which is predicted to be the fastest possible magnetization reversal mechanism.

\section*{Acknowledgements}
D.P. and S.B. acknowledges support from the European Research Council, Starting Grant 715452 ``MAGNETIC-SPEED-LIMIT''. M.H. and S.B. gratefully acknowledges support from the Swedish Research Council grant E0635001, and the Marie Sk\l{}odowska Curie Actions, Cofund, Project INCA 600398s. M.P. and P.V. acknowledge support from Basque Government under the Project n. PI2015-1-19, from MINECO under the Project FIS2015-64519-R and from the European Union under the Project H2020 FETOPEN-2016-2017 ``FEMTOTERABYTE'' (Project n. 737093). M.P. acknowledges support from MINECO through Grant BES-2013-063690 and EEBB-I-16-10873. S.U. acknowledges support from US NSF Grant Nos. ECCS-1503878 and DMR-1504449.

\section*{References}
\bibliographystyle{unsrt}

\begin{thebibliography}{10}

\bibitem{Beaurepaire1996}
E.~Beaurepaire, G.~M. Turner, S.~M. Harrel, M.~C. Beard, J.-Y. Bigot, and C.~A.
  Schmuttenmaer.
\newblock Coherent terahertz emission from ferromagnetic films excited by
  femtosecond laser pulses.
\newblock {\em Applied Physics Letters}, 84(18):3465--3467, 2004.

\bibitem{koopmans2000ultrafast}
Bert Koopmans, M.~Van~Kampen, J.~T. Kohlhepp, and W.~J.~M. De~Jonge.
\newblock Ultrafast magneto-optics in nickel: magnetism or optics?
\newblock {\em Physical Review Letters}, 85(4):844, 2000.

\bibitem{koopmans2005unifying}
Bert Koopmans, J.~J.~M. Ruigrok, Francesco Dalla~Longa, and W.~J.~M. De~Jonge.
\newblock Unifying ultrafast magnetization dynamics.
\newblock {\em Physical review letters}, 95(26):267207, 2005.

\bibitem{RasingPRB2010}
Andrei Kirilyuk, Alexey~V. Kimel, and Theo Rasing.
\newblock Ultrafast optical manipulation of magnetic order.
\newblock {\em Rev. Mod. Phys.}, 82:2731--2784, Sep 2010.

\bibitem{PhysRevB.78.174422}
E.~Carpene, E.~Mancini, C.~Dallera, M.~Brenna, E.~Puppin, and S.~De~Silvestri.
\newblock Dynamics of electron-magnon interaction and ultrafast demagnetization
  in thin iron films.
\newblock {\em Phys. Rev. B}, 78(17):174422, 11 2008.

\bibitem{Munzenberg:NatureMaterials:2010}
Markus~G. M\"{u}nzenberg.
\newblock Magnetization dynamics: Ferromagnets stirred up.
\newblock {\em Nature Materials}, 9(3):184--185, 3 2010.

\bibitem{Stamm:NatureMaterials:2007}
C.~Stamm, T.~Kachel, N.~Pontius, R.~Mitzner, T.~Quast, K.~Holldack, S.~Khan,
  C.~Lupulescu, E.~F. Aziz, M.~Wietstruk, H.~A. D\"{u}rr, and W.~Eberhardt.
\newblock Femtosecond modification of electron localization and transfer of
  angular momentum in nickel.
\newblock {\em Nature Materials}, 6(10):740--743, 8 2007.

\bibitem{dalla2007influence}
Francesco Dalla~Longa, J.~T. Kohlhepp, W.~J.~M. De~Jonge, and Bert Koopmans.
\newblock Influence of photon angular momentum on ultrafast demagnetization in
  nickel.
\newblock {\em Physical Review B}, 75(22):224431, 2007.

\bibitem{malinowski2008control}
G.~Malinowski, F.~Dalla~Longa, J.~H.~H. Rietjens, P.~V. Paluskar, R.~Huijink,
  H.~J.~M. Swagten, and B.~Koopmans.
\newblock Control of speed and efficiency of ultrafast demagnetization by
  direct transfer of spin angular momentum.
\newblock {\em Nature Physics}, 4(11):855--858, 2008.

\bibitem{Koopmans:NatureMaterials:2009}
B.~Koopmans, G.~Malinowski, F.~Dalla~Longa, D.~Steiauf, M.~F\"{a}hnle, T.~Roth,
  M.~Cinchetti, and M.~Aeschlimann.
\newblock Explaining the paradoxical diversity of ultrafast laser-induced
  demagnetization.
\newblock {\em Nature Materials}, 12 2009.

\bibitem{Boeglin:Nature:}
C.~Boeglin, E.~Beaurepaire, V.~Halt\'{e}, V.~L\'{o}pez-Flores, C.~Stamm,
  N.~Pontius, H.~A. D\"{u}rr, and J.~Y Bigot.
\newblock Distinguishing the ultrafast dynamics of spin and orbital moments in
  solids.
\newblock {\em Nature}, 465(7297):458.

\bibitem{kirilyuk2010ultrafast}
Andrei Kirilyuk, Alexey~V. Kimel, and Theo Rasing.
\newblock Ultrafast optical manipulation of magnetic order.
\newblock {\em Reviews of Modern Physics}, 82(3):2731, 2010.

\bibitem{battiato2010superdiffusive}
Marco Battiato, Karel Carva, and Peter~M. Oppeneer.
\newblock Superdiffusive spin transport as a mechanism of ultrafast
  demagnetization.
\newblock {\em Physical review letters}, 105(2):027203, 2010.

\bibitem{radu2011transient}
I.~Radu, K.~Vahaplar, C.~Stamm, T.~Kachel, N.~Pontius, H.~A. D\"{u}rr, T.~A.
  Ostler, J.~Barker, R.~F.~L. Evans, and R.~W. Chantrell.
\newblock Transient ferromagnetic-like state mediating ultrafast reversal of
  antiferromagnetically coupled spins.
\newblock {\em Nature}, 472(7342):205--208, 2011.

\bibitem{mathias2012probing}
Stefan Mathias, La-O .~O. Chan, Patrik Grychtol, Patrick Granitzka, Emrah
  Turgut, Justin~M. Shaw, Roman Adam, Hans~T. Nembach, Mark~E. Siemens, and
  Steffen Eich.
\newblock Probing the timescale of the exchange interaction in a ferromagnetic
  alloy.
\newblock {\em Proceedings of the National Academy of Sciences},
  109(13):4792--4797, 2012.

\bibitem{guidoni2002magneto}
Luca Guidoni, Eric Beaurepaire, and Jean-Yves .~Y. Bigot.
\newblock Magneto-optics in the ultrafast regime: Thermalization of spin
  populations in ferromagnetic films.
\newblock {\em Physical review letters}, 89(1):017401, 2002.

\bibitem{ostler2012ultrafast}
T.~A. Ostler, J.~Barker, R.~F.~L. Evans, R.~W. Chantrell, U.~Atxitia,
  O.~Chubykalo-Fesenko, S.~El~Moussaoui, LBPJ Le~Guyader, E.~Mengotti, and
  L.~J. Heyderman.
\newblock Ultrafast heating as a sufficient stimulus for magnetization reversal
  in a ferrimagnet.
\newblock {\em Nature communications}, 3:666, 2012.

\bibitem{mentink2012ultrafast}
J.~H. Mentink, Johan Hellsvik, D.~V. Afanasiev, B.~A. Ivanov, A.~Kirilyuk,
  A.~V. Kimel, Olle Eriksson, M.~I. Katsnelson, and Th~Rasing.
\newblock Ultrafast spin dynamics in multisublattice magnets.
\newblock {\em Physical review letters}, 108(5):057202, 2012.

\bibitem{schellekens2013comparing}
A.~J. Schellekens and B.~Koopmans.
\newblock Comparing ultrafast demagnetization rates between competing models
  for finite temperature magnetism.
\newblock {\em Physical review letters}, 110(21):217204, 2013.

\bibitem{carva2013ab}
Karel Carva, Marco Battiato, Dominik Legut, and Peter~M. Oppeneer.
\newblock Ab initio theory of electron-phonon mediated ultrafast spin
  relaxation of laser-excited hot electrons in transition-metal ferromagnets.
\newblock {\em Physical Review B}, 87(18):184425, 2013.

\bibitem{Turgut:PhysicalReviewLetters:2013}
Emrah Turgut, Chan La-o vorakiat, Justin~M. Shaw, Patrik Grychtol, Hans~T.
  Nembach, Dennis Rudolf, Roman Adam, Martin Aeschlimann, Claus~M. Schneider,
  Thomas~J. Silva, Margaret~M. Murnane, Henry~C. Kapteyn, and Stefan Mathias.
\newblock Controlling the competition between optically induced ultrafast
  spin-flip scattering and spin transport in magnetic multilayers.
\newblock {\em Physical Review Letters}, 110(19), 5 2013.

\bibitem{bergeard2014ultrafast}
N.~Bergeard, V.~L\'{o}pez-Flores, V.~Halt\'{e}, M.~Hehn, C.~Stamm, N.~Pontius,
  E.~Beaurepaire, and C.~Boeglin.
\newblock Ultrafast angular momentum transfer in multisublattice ferrimagnets.
\newblock {\em Nature communications}, 5, 3 2014.

\bibitem{Mangin:NatureMaterials:2014}
S.~Mangin, M.~Gottwald, C-H .~H. Lambert, D.~Steil, V.~Uhl\'{i}, L.~Pang,
  M.~Hehn, S.~Alebrand, M.~Cinchetti, G.~Malinowski, Y.~Fainman,
  M.~Aeschlimann, and E.~E. Fullerton.
\newblock Engineered materials for all-optical helicity-dependent magnetic
  switching.
\newblock {\em Nature Materials}, 13(3):286, 2 2014.

\bibitem{lambert2014all}
Charles-Henri .~H. Lambert, Stephane Mangin, B.~S. D. Ch~S. Varaprasad, Y.~K.
  Takahashi, M.~Hehn, M.~Cinchetti, G.~Malinowski, K.~Hono, Y.~Fainman, and
  M.~Aeschlimann.
\newblock All-optical control of ferromagnetic thin films and nanostructures.
\newblock {\em Science}, 345(6202):1337--1340, 2014.

\bibitem{StefanoPRL2016}
S.~Bonetti, M.~C. Hoffmann, M.-J. Sher, Z.~Chen, S.-H. Yang, M.~G. Samant,
  S.~S.~P. Parkin, and H.~A. D\"urr.
\newblock Thz-driven ultrafast spin-lattice scattering in amorphous metallic
  ferromagnets.
\newblock {\em Phys. Rev. Lett.}, 117:087205, Aug 2016.

\bibitem{Hauri2016}
Mostafa Shalaby, Carlo Vicario, and Christoph~P. Hauri.
\newblock Low frequency terahertz-induced demagnetization in ferromagnetic
  nickel.
\newblock {\em Applied Physics Letters}, 108(18):182903, 2016.

\bibitem{shalaby2016simultaneous}
Mostafa Shalaby, Carlo Vicario, and Christoph~P. Hauri.
\newblock Simultaneous electronic and the magnetic excitation of a ferromagnet
  by intense thz pulses.
\newblock {\em New Journal of Physics}, 18(1):013019, 2016.

\bibitem{rhie2003femtosecond}
H-S Rhie, HA~D{\"u}rr, and W~Eberhardt.
\newblock Femtosecond electron and spin dynamics in n i/w (110) films.
\newblock {\em Physical review letters}, 90(24):247201, 2003.

\bibitem{Hauri2013nature}
Carlo Vicario, Clemens Ruchert, Fernando Ardana-Lamas, Peter~M Derlet, B~Tudu,
  Jan Luning, and Christoph~P Hauri.
\newblock Off-resonant magnetization dynamics phase-locked to an intense
  phase-stable terahertz transient.
\newblock {\em Nature Photonics}, 7(9):720, 2013.

\bibitem{KentSpintronics2010}
{Andrew D.} Kent.
\newblock Spintronics: Perpendicular all the way.
\newblock {\em Nature Materials}, 9(9):699--700, 9 2010.

\bibitem{PolleyPMATHz2015}
Semanti Pal, Debanjan Polley, Rajib~Kumar Mitra, and Anjan Barman.
\newblock Ultrafast dynamics and thz oscillation in [co/pd]8 multilayers.
\newblock {\em Solid State Communications}, 221:50 -- 54, 2015.

\bibitem{stohr2007magnetism}
Joachim St{\"o}hr and Hans~Christoph Siegmann.
\newblock {\em Magnetism: from fundamentals to nanoscale dynamics}, volume 152.
\newblock Springer Science \& Business Media, 2007.

\bibitem{analyticaldamping2003}
G.~Bertotti, I.~Mayergoyz, C.~Serpico, and M.~Dimian.
\newblock Comparison of analytical solutions of landau–lifshitz equation for
  “damping” and “precessional” switchings.
\newblock {\em Journal of Applied Physics}, 93(10):6811--6813, 2003.

\bibitem{OH12008}
Fabian D.~J. Brunner, O-Pil Kwon, Seong-Ji Kwon, Mojca Jazbin\v{s}ek, Arno
  Schneider, and Peter G\"{u}nter.
\newblock A hydrogen-bonded organic nonlinear optical crystal for
  high-efficiency terahertz generation and detection.
\newblock {\em Opt. Express}, 16(21):16496--16508, Oct 2008.

\bibitem{hoffmann2011coherent}
Matthias~C. Hoffmann, Sebastian Schulz, Stephan Wesch, Steffen Wunderlich,
  Andrea Cavalleri, and Bernhard Schmidt.
\newblock Coherent single-cycle pulses with mv/cm field strengths from a
  relativistic transition radiation light source.
\newblock {\em Optics letters}, 36(23):4473--4475, 2011.

\bibitem{chen2008THzactiveMM}
Hou-Tong Chen, Willie~J Padilla, Richard~D Averitt, Arthur~C Gossard, Clark
  Highstrete, Mark Lee, John~F O’Hara, and Antoinette~J Taylor.
\newblock Electromagnetic metamaterials for terahertz applications.
\newblock {\em Terahertz Science and Technology}, 1(1):42--50, 2008.

\bibitem{pancaldiTHzantenna2017}
Michael Kozina, Matteo Pancaldi, Christian Bernhard, Tim van Driel, J~Michael
  Glownia, Premysl Marsik, Milan Radovic, Carlos~AF Vaz, Diling Zhu, Stefano
  Bonetti, et~al.
\newblock Local terahertz field enhancement for time-resolved x-ray
  diffraction.
\newblock {\em Applied Physics Letters}, 110(8):081106, 2017.

\bibitem{PolleyTHzanti}
Kumar Neeraj, Samiran Choudhury, Debanjan Polley, Rakhi Acharya, Jaivardhan
  Sinha, Anjan Barman, and Rajib~Kumar Mitra.
\newblock Efficient terahertz anti-reflection properties of metallic anti-dot
  structures.
\newblock {\em Optics Letters}, 42(9):1764--1767, 2017.

\bibitem{AuTHzValue}
Qi-Ye Wen, Huai-Wu Zhang, Yun-Song Xie, Qing-Hui Yang, and Ying-Li Liu.
\newblock Dual band terahertz metamaterial absorber: Design, fabrication, and
  characterization.
\newblock {\em Applied Physics Letters}, 95(24):241111, 2009.

\bibitem{thzglass}
Mira Naftaly and Robert~E Miles.
\newblock Terahertz time-domain spectroscopy for material characterization.
\newblock {\em Proceedings of the IEEE}, 95(8):1658--1665, 2007.

\bibitem{tudosa2004ultimate}
I~Tudosa, Ch~Stamm, AB~Kashuba, F~King, et~al.
\newblock The ultimate speed of magnetic switching in granular recording media.
\newblock {\em Nature}, 428(6985):831, 2004.

\bibitem{gamble2009electric}
SJ~Gamble, Mark~H Burkhardt, A~Kashuba, Rolf Allenspach, Stuart~SP Parkin,
  HC~Siegmann, and J~St{\"o}hr.
\newblock Electric field induced magnetic anisotropy in a ferromagnet.
\newblock {\em Physical review letters}, 102(21):217201, 2009.

\end{thebibliography}

\end{document}